\documentclass[pra,showpacs,showkeys,twocolumn]{revtex4}
\usepackage{graphicx,amsmath,amsfonts,amssymb}
\usepackage{times}
\begin{document}
\title{\Large Optimal ancilla-free phase-covariant telecloning of qudits via nonmaximally entangled states}
\author{Xin-Wen Wang$^{1,}$\footnote{E-mail: xwwang@mail.bnu.edu.cn} and Guo-Jian Yang$^{2}$}
 \affiliation{$^1$Department of Physics and Electronic Information,
  Hengyang Normal University, Hengyang 421008, People's Republic of China\\
  $^2$Department of Physics, Beijing Normal University, Beijing 100875, People's Republic of China}

\begin{abstract}
We study the $1\rightarrow 2$ phase-covariant telecloning of a qudit
without ancilla. We show that the fidelity of the two clones can
reach that of the clones in the optimal ancilla-based $1\rightarrow
2$ phase-covariant cloning and telecloning, i.e., the limitation of
quantum mechanics. More interestingly, it is a nonmaximally
entangled state rather than the maximally entangled state that can
be used to realize such a telecloning task.

\end{abstract}

\pacs{03.67.-a, 03.65.Ud}

\keywords{Ancilla-free phase-covariant telecloning, qudit, fidelity,
entanglement}

\maketitle

It is impossible to exactly copy (that is, clone) an arbitrary
quantum state because of the linearity of quantum mechanics
\cite{299N802,92PLA271}. Nevertheless, the question of how well one
can clone an unknown or partially unknown quantum state has been
attracting much interest \cite{77RMP1225} since Bu\v{z}ek and
Hillery \cite{54PRA1844} first introduced the concept of approximate
quantum copying , because it is closely related to quantum
computation, quantum communication, and quantum cryptography (see,
e.g., \cite{62PRA022301,228PLA13,88PRL127901,73PRA032304}), and can
also reveal some peculiar entanglement properties (see, e.g.,
\cite{81PRA032323,79PRA062315,27CPL100303}). If the input quantum
state is chosen from a subset of linear independent states, exact
copying can be realized probabilistically
\cite{80PRL4999,83PRL2849}. For the input state
$|\psi\rangle=\sum_{j=0}^{d-1}\alpha_j e^{i\theta_j}|j\rangle$
($d\geq 2$ is the dimension) with $\alpha_j$ being real numbers
satisfying the normalization condition
$\sum_{j=0}^{d-1}\alpha^2_j=1$ and $\theta_j\in[0,2\pi)$, three
types of (approximate) quantum cloning have been intensively
studied, i.e., universal quantum cloning with $\alpha_j$ and
$\theta_j$ being completely unknown
\cite{79PRL2153,58PRA1827,81PRL5003}, real state cloning with
$\theta_j=0$ and $\alpha_j$ being unknown
\cite{62PRA012302,68PRA032313,75PRA044303}, and phase-covariant
cloning with $\alpha_j=1/\sqrt{d}$ and $\theta_j$ being unknown
\cite{62PRA012302,67PRA022317}. In general, the more the information
about the input state is known, the better the state can be cloned.
As a consequence, the optimal fidelities of clones (the fidelity
limit that quantum mechanics allows) in the real state cloning and
phase-covariant cloning are higher than that in the universal
quantum cloning. Recently, more attention was paid to
phase-covariant cloning because of its use in connection with
quantum cryptography \cite{67PRA012311}.

Quantum cloning process can be regarded as distribution of quantum
information from the initial system to a larger one. Thus quantum
cloning combining with other quantum information processing tasks
may have potential applications in quantum communication,
distributed quantum computation, and so on
\cite{63PRA042308,95PRL090504}. This leads to the advent of the
concept of telecloning \cite{57PRA2368}, which is the combination of
quantum cloning and quantum teleportation \cite{70PRL1895}.
Telecloning functions as transmitting multiple copies of an unknown
(or partially unknown) quantum state to distant sites, i.e.,
realizing one-to-many nonlocal cloning, via previously shared
multipartite entangled states. The entanglement channel for
telecloning can be directly constructed by the corresponding cloning
transformation \cite{59PRA156}.

In the aforementioned quantum cloning and telecloning, the ancillas
(extra quantum systems besides the ones used to carry the cloned
states) plays an important role. Recently, quantum cloning without
ancillas, i.e., the so-called ancilla-free (or economical) cloning
\cite{60PRA2764,69PRA062316,71PRA042327,72PRA052322}, has attracted
much interest, because it may be easier than the one with ancillas
for experimental implementation \cite{94PRL04005}. Durt \emph{et
al.} showed that an ancilla-free version of the $1\rightarrow2$
universal cloning with the optimal fidelity (the fidelity limit that
quantum mechanics allows) cannot be realized in any dimension, and
ancilla-free versions of both the $1\rightarrow2$ Fourier-covariant
\cite{84PRL4497} and phase-covariant cloning with the optimal
fidelity can be implemented only for qubits. They also presented an
ancilla-free phase-covariant cloning machine for qudits, with the
fidelity being lower than that of the optimal phase-covariant
cloning machine involving an ancilla. Because of the relationship
between cloning and the corresponding telecloning
\cite{59PRA156,67PRA012323,369PLA112}, their conclusions also imply
that the ancilla-free $1\rightarrow2$ phase-covariant telecloning
with the optimal fidelity for qudits and universal telecloning with
the optimal fidelity in any dimension cannot be realized via the
maximally entangled states constructed by the corresponding cloning
transformations.

In this letter, we present a scheme for ancilla-free $1\rightarrow
2$ phase-covariant telecloning of a qudit. We show that the fidelity
can reach that of the $1\rightarrow 2$ phase-covariant cloning
machine of Ref.~\cite{67PRA022317} and telecloning machine of
\cite{67PRA012323}. That is, the fidelity of the clones in our
ancilla-free telecloning scheme can hit to the optimal fidelity that
quantum mechanics allowed. More interestingly, the suitable quantum
channel for realizing the above telecloning task is a nonmaximally
entangled state rather than the maximally entangled state.

 First, we briefly review
Durt's ancilla-free $1\rightarrow2$ phase-covariant (symmetric)
cloning machine for a $d$-dimensional system. For the input state
\begin{equation}
  |\psi^{in}\rangle_1=\frac{1}{\sqrt{d}}\sum\limits_{j=0}^{d-1}e^{i\theta_j}|j_1\rangle,
\end{equation}
the cloning machine (transformation) functions as \cite{72PRA052322}
\begin{eqnarray}
|j_10_2\rangle\rightarrow |\phi_{j}^{0}\rangle_{12}
 \end{eqnarray}
where
\begin{eqnarray}
 && |\phi^{k}_{k}\rangle_{12}=|k_1k_2\rangle\nonumber\\
 && |\phi^{k}_{j}\rangle_{12}=\frac{1}{\sqrt{2}}(|j_1k_2\rangle+|k_1j_2\rangle),~j\neq k.
\end{eqnarray}
Here, we have assumed that the second quantum system (carrier) is
initially in the state $|0_2\rangle$. The output state reads
\begin{equation}
\label{out0}
|\psi^{out}\rangle_{12}=\frac{1}{\sqrt{d}}\sum\limits_{j=0}^{d-1}e^{i\theta_j}|\phi^0_{j}\rangle_{12}.
\end{equation}
The fidelity of each clone (copy) is
\begin{eqnarray}
\label{F}
 F_{econ}(d)&=&\langle\psi^{in}|_{1(2)}\mathrm{Tr}_{2(1)}\left(|\psi^{out}\rangle_{12}\langle\psi^{out}|\right)|\psi^{in}\rangle_{1(2)}\nonumber\\
  &=& \frac{1}{2d^2}\left[(d-1)^2+(1+2\sqrt{2})(d-1)+2\right].
\end{eqnarray}
However, the optimal fidelity of $1\rightarrow2$ phase-covariant
cloning (with an ancilla) is \cite{67PRA022317}
\begin{equation}
  F_{opt}(d)=\frac{1}{4d}\left(d+2+\sqrt{d^2+4d-4}\right).
\end{equation}
It can be verified that for $d=2$, $F_{econ}(2)=F_{opt}(2)$, while
 for $d>2$, $F_{econ}(d)<F_{opt}(d)$. Thus this type of
ancilla-free phase-covariant cloning is ``suboptimal''.

We now describe our telecloning protocol. The task is: Alice wants
to transmit one copy of the state $|\psi^{in}\rangle_{A_1}$ of
particle $A_1$ to distant Bob and Charlie, respectively. Assume that
the quantum channel among them is a three-particle entangled state
as follows:
\begin{equation}
\label{channel}
 |\Psi\rangle_{A_2BC}=\sum\limits_{j=0}^{d-1}x_j|j_{A_2}\rangle|\phi^0_{j}\rangle_{BC},
\end{equation}
where $x_j$ are probability amplitudes satisfying normalization
condition $\sum_{j=0}^{d-1}x_j^2=1$. For simplicity, we have assumed
that $x_j$ are real numbers. Here, particle $A_2$ is on Alice's
hand, and particles $B$ and $C$ are held by Bob and Charlie,
respectively. The von Neumann entropy of
$\rho_{A_2}=\mathrm{tr}_{BC}(|\Psi\rangle_{A_2BC}\langle\Psi |)$ is
\begin{equation}
\label{entropy}
  S(\rho_{A_2})=-\sum\limits_{j=0}^{d-1}x^2_j\log_2x^2_j.
\end{equation}
The state of the total system is
\begin{eqnarray}
|\Psi\rangle_{total}&=&|\psi^{in}\rangle_{A_1}\otimes|\Psi\rangle_{A_2BC}\nonumber\\
&=&\frac{1}{d}\sum\limits_{l=0}^{d-1}\sum\limits_{k=0}^{d-1}|\Phi\rangle^{lk}_{A_1A_2}\nonumber\\
 &&  \times\sum\limits_{j=0}^{d-1}e^{-2\pi ijk/d}x_{j\oplus l} e^{i\theta_j}|\phi^0_{j\oplus
   l}\rangle_{BC},
\end{eqnarray}
where $j\oplus l$ denotes $j+l$ modulo $d$ and
$|\Phi\rangle^{lk}_{A_1A_2}$ are generalized Bell-basis states given
by
\begin{equation}
  |\Phi\rangle^{lk}_{A_1A_2}=\frac{1}{\sqrt{d}}\sum\limits_{j=0}^{d-1}\exp\left(\frac{2\pi
  ijk}{d}\right)|j\rangle|j\oplus l\rangle.
\end{equation}
In order to realize the telecloning task, Alice performs a complete
projective measurement jointly on particles $A_1$ and $A_2$ in the
generalized Bell-basis
$\{|\Phi\rangle^{lk}_{A_1A_2},~l,k=0,1,2\cdots,d-1\}$, and informs
Bobs of the outcome.

If Alice gets the outcomes $|\Phi\rangle^{0k}_{A_1A_2}$, the state
of particles $B$ and $C$ collapses into
\begin{equation}
\label{out1}
   |\psi\rangle_{BC}=\sum\limits_{j=0}^{d-1}e^{-2\pi ijk/d}x_j e^{i\theta_j}|\phi^0_{j}\rangle_{BC}.
\end{equation}
After receiving the measurement outcome, Bob and Charlie perform,
respectively, their particles the following local operation:
\begin{equation}
   U^{0k}_{B(C)}=\sum\limits_{j=0}^{d-1}\exp\left(\frac{2\pi ijk}{d}\right)|j\rangle _{B(C)}\langle j|.
\end{equation}
Then the state of Eq.~(\ref{out1}) evolves into
\begin{equation}
 |\psi^{out'}\rangle_{BC}= \sum\limits_{j=0}^{d-1}x_j e^{i\theta_j}|\phi^0_{j}\rangle_{BC}.
\end{equation}
The fidelity of clones that Bob and Charlie obtained is
\begin{equation}
\label{fidelity}
 F^t_{econ}(d)=\frac{1}{d}\left(1+\sqrt{2}x_0\sum\limits_{j=1}^{d-1}x_j+\sum\limits_{j=1}^{d-2}\sum\limits_{k=j+1}^{d-1}x_jx_k\right).
\end{equation}

If Alice gets the outcome $|\Phi\rangle^{lk}_{A_1A_2}$ ($l\neq 0$),
the state of particles $\{1,2,\cdots,n\}$ collapses into
\begin{eqnarray}
\label{out2}
 |\tilde{\psi}\rangle_{BC}
  =\sum\limits_{j=0}^{d-1}e^{-i2\pi jk/d}x_{j\oplus l} e^{i\theta_j}|\phi_{j\oplus l}^{0}\rangle_{BC}.
\end{eqnarray}
After receiving the measurement outcome, Bobs perform, respectively,
their particles the following local operation:
\begin{eqnarray}
 U^{lk}_{B(C)}=\sum\limits_{j=0}^{d-1}\exp\left(i\frac{2\pi jk}{d}\right)|j\rangle_{B(C)}\langle j\oplus l|.
\end{eqnarray}
The state of Eq.~(\ref{out2}) evolves into
\begin{eqnarray}
\label{fout2}
 |\psi\rangle^{out''}_{BC}=\sum\limits_{j=0}^{d-1}x_{j\oplus l} e^{i\theta_j}|\phi_{j}^{d-l}\rangle_{BC},
\end{eqnarray}
where a nonsense global phase factor $\exp[i2\pi(d-l)k/d]$ is
discarded. It can be easily verified that $\{x_{j\oplus
l}|\phi_{j}^{d-l}\rangle_{BC},~j=0,1,\cdots,d-1\}$ have the same
contributions with $\{x_j|\phi_{j}^0\rangle_{BC}\}$ to the cloning
fidelity. Thus the fidelity of each clone is also equal to $
F^t_{econ}(d)$ in this case. Unlike Ref.~\cite{59PRA156}, our
protocol does not involve ancilla and thus it is ancilla-free.

For the case $x_j=1/\sqrt{d}$, $S(\rho_{A_2})=\log_2 d$ and the
quantum channel is a maximally entangled state in terms of the
subsystem of Alice (particle $A_2$) and the subsystem of Bob and
Charlie (particles $B$ and $C$). Then $F^t_{econ}(d)=F_{econ}(d)$
less than $F_{opt}(d)$ for $d>2$. In the following, we shall show
that the fidelity $F^t_{econ}(d)$ can be equal to $ F_{opt}(d)$ for
any $d$ with another choice of $\{x_j\}$.

We set
\begin{eqnarray}
\label{XY}
  && x_0=X(d)=\sqrt{\frac{4(d-1)}{D(D+d-2)}},\nonumber\\
  && x_j=Y(d)=\sqrt{\frac{d^2+(d-2)D}{D(D+d-2)(d-1)}},~ j\neq 0,
\end{eqnarray}
where $D=\sqrt{d^2+4d-4}$. Then it can be verified that
$F^t_{econ}(d)= F_{opt}(d)$ for any $d$. In fact, the output state
($|\psi^{out'}\rangle_{BC}\langle\psi^{out'}|$) of our telecloner is
then equivalent to that ($\rho^{out}_{opt}$) of the optimal
phase-covariant cloner after tracing out the ancilla
\cite{67PRA022317,67PRA012311}. Particularly,
$\rho^{out}_{opt}=|\psi^{out'}\rangle_{BC}\langle\psi^{out'}|+\tilde{\rho}$
with
$\langle\psi^{in}|_{B(A)}\mathrm{tr}_{A(B)}(\tilde{\rho})|\psi^{in}\rangle_{B(A)}=0$.
In this case, the entanglement channel of Eq.~(\ref{channel})
reduces to
\begin{equation}
\label{channel1}
|\Psi'\rangle_{A_2BC}=X(d)|0_{A_2}\rangle|\phi^{0}_0\rangle_{BC}+Y(d)\sum\limits_{j=1}^{d-1}|j_{A_2}\rangle|\phi_{j}^0\rangle_{BC}.
\end{equation}
When $d=2$, $S(\rho_{A_2})=1$ and $|\Psi'\rangle_{A_2BC}$ is a
maximally entangled state. For $d>2$, however, the amount of
entanglement with von Neumann measure between particle $A_2$ and
particles (B,C) is $E(|\Psi'\rangle_{A_2(BC)})=-X^2\log_2
X^2-(d-1)Y^2\log_2 Y^2<\log_2 d$, which implies that the subsystem
of Alice (sender) and the subsystem of Bob and Charlie (receivers)
in the state of Eq.~(\ref{channel1}) are only partially entangled.
Thus we can safely conclude that the ancilla-free $1\rightarrow 2$
phase-covariant telecloning with the optimal fidelity for qudits can
be realized via suitable nonmaximally entangled states acting as the
quantum channel.

In order to reveal clearly the relationship between the fidelity of
clones and the amount of entanglement of the quantum channel, we
show how $F^t_{econ}(d)$ varies with the variation of von Neumann
entropy $S(\rho_{A_2})$ in Fig.~1. For simplicity, we have assumed
that $x_1=x_2=\cdots=x_{d-1}$. It can be seen that for $d=2$, the
increase (decrease) in $S(\rho_{A_2})$ always leads to increase
(decrease) in $F^t_{econ}(2)$. For $d>2$, however, a
counterintuitive phenomenon appears: when $1/\sqrt{d}\leq x_0\leq
X(d)$, $F^t_{econ}(d)$ increases (decreases) with the decrease
(increase) in $S(\rho_{A_2})$.

In conclusion, we have studied the ancilla-free $1\rightarrow 2$
phase-covariant telecloning for qudits. We have shown that the
fidelity can reach that of the $1\rightarrow 2$ phase-covariant
cloning machine of Ref.~\cite{67PRA022317}. In other words, the
fidelity of the clones in our ancilla-free teleclonig scheme can hit
to the optimal fidelity (the fidelity limit that quantum mechanics
allows for phase-covariant cloning). We have also shown that the
increase (decrease) in amount of entanglement of the quantum channel
may lead to the decrease (increase) in the fidelity of clones in the
ancilla-free phase-covariant telecloning for qudits. This effect
leads to another interesting phenomenon: the suitable quantum
channels for realizing the optimal ancilla-free $1\rightarrow 2$
phase-covariant telecloning of qudits are special configurations of
nonmaximally entangled states rather than the maximally entangled
states. Note that nonmaximally entangled states can be better than
the maximally entangled states for several other quantum tasks has
also been reported \cite{100PRL110503}.

\begin{acknowledgements}
This research was supported by National Natural Science Foundation
of China (Grant No. 11004050).
\end{acknowledgements}

\begin{widetext}
\begin{figure}[htp]
  \includegraphics[width=12cm,height=10cm]{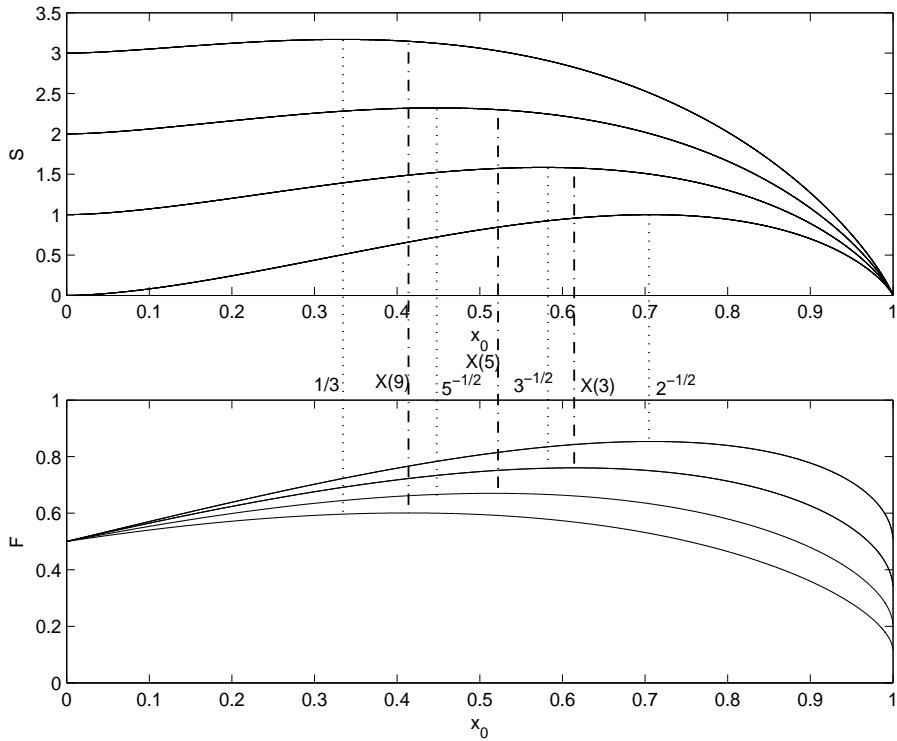}
\caption{The von Neumann entropy $S(\rho_{A_2})$ (upper graph) and
the fidelity $F^t_{econ}(d)$ (lower graph) versus the probability
amplitude $x_0$, where $x_1=x_2=\cdots=x_{d-1}$. From bottom (top)
to top (bottom) in the upper (lower) graph, the curves correspond to
$d=2,3,5,$ and 9, respectively. The vertical dotted lines ending in
the corresponding curves represent that $S$ reaches the maximum when
$x_0=1/\sqrt{d}$, and the dashdotted lines denote that $F$ hits to
the maximum when $x_0=X(d)$.} \label{figure}
\end{figure}
\end{widetext}
\end{document}